\documentclass[twocolumn,prl,aps,floatfix]{revtex4} 
\usepackage{times}
\usepackage{graphicx}
\usepackage{dcolumn}
\usepackage{bm}

\begin{document}

\preprint{APS/123-QED}

\title{ The pseudogap in YBa$_2$Cu$_3$O$_{6+\delta}$ is not bounded by a line of phase transitions - thermodynamic evidence }

\author{J. R. Cooper$^{1}$, J.W. Loram$^{1}$, I. Kokanovi\'{c}$^{1,2}$,  J. G. Storey$^{3}$ and J. L. Tallon$^{3}$}

\affiliation{$^1$Cavendish Laboratory, Cambridge University, CB3 0HE, United Kingdom.
$^2$Department of Physics, Faculty of Science, University of Zagreb,
P.O.Box 331, Zagreb, Croatia.\\
$^3$MacDiarmid Institute and Robinson Research Institute, Victoria University of
Wellington, P.O. Box 31310, Lower Hutt, New Zealand.}

\date{\today}

\begin{abstract}

 We discuss a recent resonant ultrasound spectroscopy (RUS) study of
YBa$_2$Cu$_3$O$_{6+\delta}$, which infers a line of phase transitions bounding the
pseudogap phase and argue that this scenario is not supported by thermodynamic
evidence. We show that the  anomalies in RUS, heat capacity and thermal expansion at
the superconducting transition temperatures agree  well.  But there are large
discrepancies between RUS and thermodynamic measurements at  $T^*$ where the pseudogap
phase transitions are purported to occur. Moreover, the frequency and temperature
dependence of the RUS data for the crystal with $\delta = 0.98$, interpreted in terms
of critical slowing down near an electronic phase transition, is five orders of
magnitude smaller than what is expected. For this crystal  the RUS data near $T^*$ are
more consistent with non-equilibrium effects such as oxygen relaxation.
\end{abstract}


 \maketitle

Hole-doped high-$T_c$ superconducting (SC) cuprates have a partial gap in the
electronic density of states (DOS), above and below $T_c$, in both under- and
optimally doped regions of their phase diagram \cite{Loram93,Loram01}. This partial
gap is referred to as a ``pseudogap" because it's spectroscopic signatures involve
loss of spectral weight near
 the Fermi level \cite{Timusk,TallonLoram}. For many years debate focused on whether the pseudogap is a precursor or
  competitor to superconductivity \cite{Norman}. More recently there has been a consolidation of support for the latter
  view - the pseudogap coexists with superconductivity and depletes spectral weight otherwise available for
  superconductivity \cite{Bernhard,Vishik,Taillefer,Sacuto}.
  As a consequence ground-state properties
   such as superfluid density, quasiparticle weight, critical fields and condensation energy become sharply
   diminished as $p$, the number of doped holes per Cu, falls below the critical doping, $p_{crit}$, at which
    the pseudogap opens \cite{Loram01,TallonLoram}. This ground-state crossover from strong superconductivity
    in the over-doped region to weak superconductivity in the coexisting pseudogap region is remarkably
    abrupt \cite{Loram01,Anukool}.

    This raises the further critical question as to whether this abrupt
crossover is in fact a phase transition. Is the pseudogap state a thermodynamic phase
bounded by a line of phase transitions? We have consistently argued that
 it is not - no specific heat anomaly at the boundary has been observed despite an
  intensive search \cite{Loram93,Loram01,Junod}.
     However ,
     there is evidence for anomalies in other properties, for
     example
     He {\it et al} \cite{Shen}
    presented a combined study of angle-resolved photoemission spectroscopy, polar Kerr effect, and
    time-resolved reflectivity, all consistent with a mean-field-like vanishing of an order parameter
    at $T^*$. With hindsight it is probable \cite{Forgan2} that many of these effects arise from the gradual onset of charge density
     wave (CDW) order in the pseudogap state.  Hard x-ray measurements of
     YBa$_2$Cu$_3$O$_{6+\delta}$(YBCO) show that this sets in  below 150~K for
     both the ortho-VIII ($\delta=0.67$) \cite{Forgan1}  and ortho-II ($\delta=0.54$)
     phases \cite{Forgan2}.
    Recently, Shekhter et al. \cite{Shekhter} conclude  from resonant ultrasound spectroscopy (RUS) studies
     that $T^*(p)$ represents a line of phase transitions, and ascribe this to the onset of a pseudogap.
     We
     discuss their work
      in detail and show that the results for
      their over-doped
      crystal are not
       consistent with changes in equilibrium thermodynamics as might be found in the neighborhood of a phase transition
        but are more consistent with non-equilibrium anelastic relaxation effects.

The conclusions advanced by Shekhter {\it et al} were based on measurements of mode
frequency, $f$, and resonance
 width, $\Gamma$, of various mode vibrations for two de-twinned single-crystal samples of
 YBCO. One
  was fully oxygenated and slightly over-doped with $T_c$ = 88~K and the other was under-doped with $T_c$ = 61.6 K. For
  the latter sample they find a change in slope $df/dT$ near 245~K, close to the doping dependent $T^*(p)$ found in neutron
  scattering \cite{Fauque} where evidence for the onset of weak magnetic order
   was reported.
   Shekhter {\it et al} also report broad peaks in $\Gamma$ at somewhat
   higher temperatures for two modes and at a lower temperature for a third mode. More dramatic effects are observed
    for the fully-oxygenated, over-doped sample, with $T_c$ = 88~K. A clear break in slope in mode frequency is observed
     at $T^*$ = 68~K as reproduced below in Fig. 1(a). Further, as shown in Fig. 1(b) a strong peak in $\Gamma$ is found
      at a slightly higher temperature which increases linearly with mode frequency. From these data
       Shekhter {\it et al.} infer the occurrence of a thermodynamic phase transition at the onset of the pseudogap at
        $T^*$. They further conclude that $T^*$ falls to zero (possibly at a quantum critical point) within the SC
        dome. Their conclusions are endorsed in a commentary by Zaanen \cite{Zaanen} who asserts that these results
        provide evidence for the current-loop model of the pseudogap due to Varma \cite{Varma}.

The fact that $T^*(p) \rightarrow 0$ within the SC dome is a feature we have noted for
15 years \cite{Loram01,TallonLoram}, though our investigations suggest that this
occurs at a significantly lower hole doping than the value in  Ref.~\cite{Shekhter} .
We first address the claim \cite{Shekhter} that the prominent RUS anomaly at 68~K in
the overdoped crystal is caused by the onset of a pseudogap.  Thermodynamic evidence
suggests
 that there is no pseudogap in fully-oxygenated YBCO. This is based on two  observations:
 (i) if as claimed in Ref. \cite{Shekhter} a pseudogap opens at $T^* \approx 68$~K which is
only a little
  below $T_c$ ($T^* \approx 0.76 T_c$)
 it would cause a large additional entropy loss. This would inevitably result in a specific heat anomaly at $T^*$ that
 is a significant fraction of the one at $T_c$, contrary to observation (see Fig 1 (a)); and (ii) key thermodynamic
features of a pseudogap ground state are a strongly reduced SC condensation energy,
superfluid density and associated critical fields \cite{Loram01} which all result from
the partial gapping of the Fermi surface by the competing pseudogap. In YBCO these
features are only present at lower hole doping, and at full oxygenation their full
values are restored \cite{Loram01,Taillefer}. In other words $T^*$ is  zero at full
oxygenation, contrary to the findings of Shekhter {\it et al}.

\begin{figure}
\centerline{\includegraphics*[width=85mm]{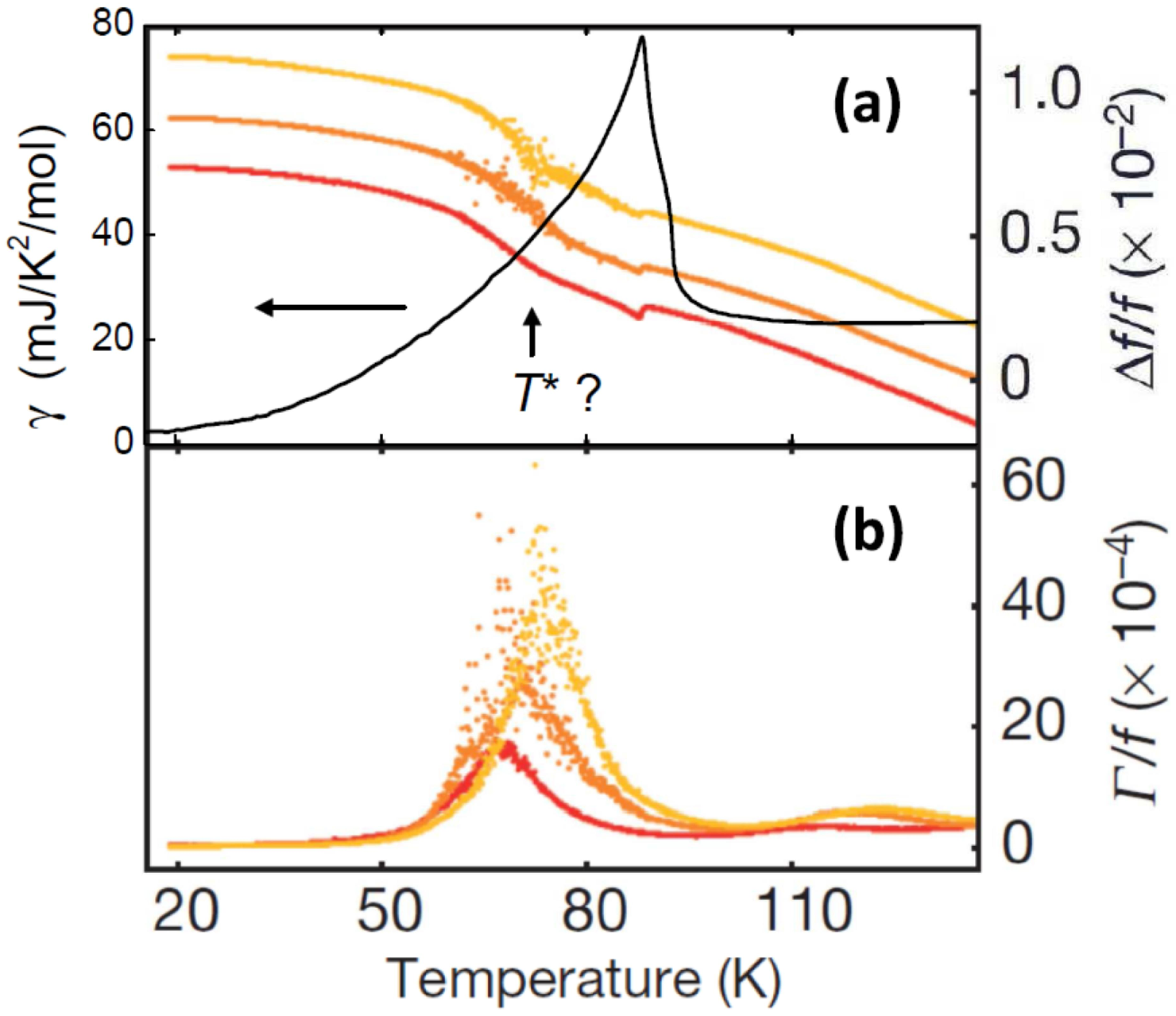}} \caption{\small (Color online)
Adaptation of  Figs. 2b and 2d from Shekhter {\it et al.} \cite{Shekhter} showing (a)
the relative change in RUS mode frequency, $\delta f/f$, for their
YBa$_2$Cu$_3$O$_{6.98}$ (YBCO) crystal superimposed on the electronic specific heat
coefficient $\gamma \equiv C_V/T$ for YBCO at full oxygenation \cite{Loram93,Loram01}.
The right-hand scale has been corrected from units of 10$^{-4}$ to 10$^{-2}$
consistent with Figs 1a and 1c of the original article and as confirmed by the authors
\cite{Shekhter2}. $T^*$ marks where a change in slope of $\delta f/f$ is observed
along with a peak in the resonance width, $\Gamma$, which is shown in (b). No feature
is observed in $\gamma(T)$ at $T^*$ even though the relative scale for $\gamma$ has
been greatly amplified so that the jump $\delta \gamma$ at $T_c$ is 90 $\times$ the
size of the jump in $\delta f/f$ at $T_c$. } \label{f&gamma}
\end{figure}

As stated in the Supplementary Material (SM) of Ref. \cite{Shekhter}, the elastic
moduli which determine the mode frequencies are related to the Helmholtz free energy
$F$ (proportional to the sample volume $V$). For an isotropic solid
 $dF = -SdT - PdV$, where $S$ is the total entropy, $T$ the absolute temperature and $PdV$
 is the work done on the solid when application of pressure $P$ causes a volume change $-dV$.
  Hence the isothermal bulk modulus
   $\kappa_T = -V \left(\partial P/\partial V\right)_T = V\left(\partial^2 F/\partial V^2\right)_T$. In the
   present case the crystal has many normal modes whose frequencies are determined by combinations of the elastic
    constants, such as $c_{11}$, $c_{12}$ and $c_{44}$ in the usual notation.  There are nine independent elastic
    constants for orthorhombic symmetry \cite{Migliori}. The crystal is a thermodynamic system and for every
    normal mode, whose frequency is measured by RUS, there will be a generalized force per unit area equivalent
     to $P$ and a corresponding deformation equivalent to $dV$ that relates the change in $F$ to the work done
      on (or by) the crystal. For example  a uniform compression in the $x$ direction gives a change in
       length $dL/L = -e_{xx}$, the force per unit area is $c_{11}e_{xx}$, the contribution
       to $F$ (the work done on the crystal)
        is $\frac{1}{2}ALc_{11}e_{xx}^2$ where $A$ is the area perpendicular to $x$
       giving
         $c_{11} = V^{-1} \left(\partial^2 F/\partial e_{xx}^2 \right)_T$.

The effect of a phase transition (i.e. the order parameter) on the elastic moduli, can
be calculated from the volume-, or more generally, the  strain-dependence of the
transition temperature $T_c$. For any second-order transition described by Landau
theory there is a jump $\delta \gamma$ in the  specific heat coefficient $\gamma
\equiv C_V/T$ at $T_c$, where $C_V =TV^{-1}\left(\partial S/\partial T\right)_V$ is
the
  specific heat capacity  at constant volume and $S =-\left(\partial F/\partial T\right)_V$.
 Integrating twice with respect to $T$ gives the decrease in
 $F$ below $T_c$ as $\delta F = -\frac{1}{2}V \delta \gamma \; (T_c-T)^2$. For an isotropic solid, differentiating
 twice with respect to volume gives the change in bulk modulus. Hence the
 fractional change in frequency at, or just below, $T_c$ for a uniform dilation mode in which the crystal
  does not change its shape is given by:

\begin{equation}
\frac{\delta f}{f} = \frac{1}{2} \frac{\delta \kappa_T}{\kappa_T} = - \frac{1}{2}
\frac{\delta \gamma\; T_c^2}{\kappa_T} \left(\frac{d(\ln T_c)}{d(\ln  V)}\right)^2 .
\label{dilation}
\end{equation}

We note  that similar arguments hold for continuous phase transitions described by
non-mean-field critical exponents, since $\left(\partial^2 F/\partial V^2 \right)_T$
will usually be dominated by a term of the form $\left(\partial ^2 F/\partial T_c^2
\right)_T\left(d T_c /d V\right)^2$.   For an isotropic classical superconductor with
$T_c= 1.14\; \theta_D \exp(-1/\lambda)$, where $\theta_D$ is the Debye temperature and
$\lambda$ the dimensionless electron-phonon coupling constant, Eq.~\ref{dilation}
leads to the formula  $\delta f/f \sim (T_c/T_F)^2$, where $T_F$ is the Fermi
temperature,  given in  the SM  of Ref. \cite{Shekhter} but with a pre-factor
$[d\ln\theta_D/d\ln V + (1/\lambda)d\ln\lambda/d\ln V]^2$. This will be of order 4 for
a strong-coupling superconductor such as  lead but much larger for a weak-coupling one
such as aluminium where $\lambda$ is small.

In the anisotropic case, for a uniform strain $e_{xx}$ along the $x$-direction and all
other strain components set to zero, Eq.~\ref{dilation} becomes

\begin{equation}
\frac{\delta f}{f} = \frac{1}{2} \frac{\delta c_{11}}{c_{11}} = - \frac{1}{2}
\frac{\delta \gamma T_c^2}{c_{11}} \left(\frac{d(\ln T_c)}{de_{xx}}\right)^2  .
 \label{strain}
\end{equation}

In general $c_{11}$ and $e_{xx}$ in Eq.~\ref{strain} should be replaced by the
appropriate linear combinations of elastic constants and strains that can be obtained
by finding the normal vibrational modes of the crystal subject to appropriate boundary
conditions \cite{Migliori}. Note that Eqs.~\ref{dilation} and~\ref{strain} always
 give a negative frequency shift in the lower $T$ phase because (i) the linear combination of elastic
  constants mentioned above must always be positive so that the restoring force
   opposes the deformation and  (ii) $\delta \gamma$ will invariably be positive because the lower $T$ phase
   will have lower entropy.

We have used $c_{11}$ as an example in Eq.~\ref{strain} but experimentally for
well-oxygenated YBCO,
 $d T_c/d e_{aa}$ = 230$\pm23$~K and $c_{aa}$ = 2310 kbar,
 while the corresponding values for the crystallographic $b$-axis are -220$\pm22$~K and 2680 kbar \cite{Welp}, assuming that
 the fractional errors are the same as those quoted for $d T_c/dP_{a,b}$.
 Simply substituting the $a$-axis values into Eq.~\ref{strain} and
 taking $\delta \gamma(T_c)$ = 56 mJ/mol/K$^2$ \cite{Loram93} or 0.54 mJ/cm$^3$/K$^2$ only
  gives $\delta f/f = 0.6 \times 10^{-4}$, over a factor of 10 smaller than the experimental value
  of 7 $\times 10^{-4}$ shown in Fig.~1c of Ref.~\cite{Shekhter}.  Although the vibrational modes are
  not specified, it is reasonable to suppose that
   the authors of Ref.~\cite{Shekhter}  showed their clearest SC anomaly for which the
   positive $a$- and negative $b$-axis $dT_c/de_{aa,bb}$ terms reinforce each other.
    Detailed calculations of the kind described in Ref.~\cite{Migliori} are clearly
    desirable. However the elastic constants given in Ref.~\cite{Welp} are
      reasonably isotropic. For isotropic cubic crystals, a standard textbook calculation
      of acoustic phonon modes \cite{Kittel} gives a transverse wave propagating in the ($1,1,0$) direction
      with a velocity equal to $\sqrt{\left(c_{11}-c_{12}\right)/(2\rho)}$ where $\rho$ is the density, for
      which the atomic displacements are along ($1,-1,0$). Generalizing Eq.~\ref{strain} to this case
      with $c_{11}$ = $c_{aa}$ and $c_{12} = c_{ab}$ = 1320 kbar \cite{Welp} and replacing $dT_c/de_{xx}$ by
       $\left(dT_c/d e_{aa}-dT_c/de_{bb}\right)/\sqrt{2}$, gives
       $\delta f/f$ = $-5.5 \pm1.5\times 10^{-4}$, in good agreement with the measured value for
       the authors' well-oxygenated YBCO crystal.

For their under-doped crystal the elastic constants are not known as precisely.
 Uniaxial pressure along the $a$-axis has a factor of 2 smaller effect on a crystal with $T_c$ = 60~K
 while the $b$-axis value is similar to that of the over-doped crystal \cite{Kraut}.
 From Fig. 1b of Ref.~\cite{Shekhter} we see that $\delta f/f$ at $T_c$ is a factor of 10 or so smaller
  for the under-doped crystal. This is consistent with the fact that $\delta \gamma(T_c)$ is a factor of 7
   smaller \cite{Loram01}, combined with the somewhat reduced value of $dT_c/de_{aa}$.

{\it We conclude that the reported anomalies in RUS at the SC transitions for both the
over-doped
 and under-doped samples are consistent with the specific heat, pressure derivatives of $T_c$  and the known values of the elastic constants}.
 In contrast to Ref.~\cite{Shekhter} we attribute the lower value of $\delta \gamma(T_c)$ for the under-doped crystal
  to the effect of the pseudogap rather than to possible oxygen disorder. (Note that quantum oscillations, which are
   extremely sensitive to disorder, have been observed in similar under-doped YBCO crystals \cite{Carrington,Sebastian}).

Turning to the RUS anomalies at the putative $T^*$ values we find a very different
picture. For both over- and under-doped samples, relatively abrupt changes in the
slope of $\delta f/f$ are observed at $T^*$, in contrast with the discontinuities seen
at the SC transitions. Interpreted in terms of a phase transition, this may perhaps be
attributed to non-mean-field critical exponents as in the  current loop model proposed
by Varma \cite{Varma}, [see comment after Eq.~\ref{dilation}]. Although
Eq.~\ref{dilation} does not strictly apply in this situation, estimates given later
suggest that the observed changes in $\delta f/f$ are incompatible with the absence of
corresponding anomalies in the specific heat.

We reproduce in Fig.~1 the $T$-dependence of $\delta f/f$ for the overdoped sample
from Fig.~2b
 of Shekhter {\it et al.} \cite{Shekhter}. Note that the right-hand scale has been corrected to
  units of $10^{-2}$ as must be the case for consistency with their Fig.~1.
  This correction in scale has been confirmed by the authors \cite{Shekhter2}.
   We superimpose on Fig.~1 the experimentally-determined electronic specific heat coefficient $\gamma(T)$
   for fully-oxygenated YBCO \cite{Loram01} where, for comparison, the scale is chosen such that the anomaly
    in $\gamma(T)$ at $T_c$ is $90\times$ the
    anomaly in $\delta f/f$ at $T_c$. The figure shows
    an abrupt increase of slope in $\delta f/f$ below $T^* \approx 68$~K.  It is more convenient  to discuss
     this increase
     and its relation with the thermal expansion coefficient $\alpha(T)$~\cite{cond_mat} in terms of the standard Ehrenfest
     equations derived via the Gibbs free energy \cite{Adkins}. Taking $dL= (\partial L/\partial T)_PdT +
     (\partial L/\partial P_x)_TdP_x$ and $dS= (\partial S/\partial T)_PdT +
     (\partial S/\partial P_x)_TdP_x$, for
     uniaxial pressure along the
     $x$ direction and requiring that there are no length ($L$) or entropy ($S$) changes at $T_c$, i.e. $dL_1=dL_2$
      and $dS_1=dS_2$ for phases 1 and 2 leads to \cite{Adkins}:

     \begin{equation}
\frac{dT_c}{dP_x} =\frac{\delta \alpha_x}{\delta\gamma}= \frac{\delta
\mathbf{S}_{xx}}{\delta\alpha_x}. \label{Gibbs}
\end{equation}

\noindent These  relate the uniaxial pressure dependence of $T_c$ to the changes in
$\gamma$ and $\alpha$ at constant pressure and the elastic constants. Here the
compliance tensor $\mathbf{S}_{ij}$ is the inverse of the full elasticity tensor
$c_{ij}$ defined earlier. Eq.~\ref{Gibbs} applies to abrupt changes, $\delta \alpha_x$
etc., at a second order phase transition and to positive or negative changes in slope
$\delta( d \alpha_x/dT)$ etc. at a third order transition. We can only invert $c_{ij}$
approximately, see footnote \onlinecite{inversion}.  Using $dT_c/dP_a=-0.2 \pm 0.02$
K/kbar~\cite{Welp}, and the same value of $\delta\gamma$, Eq.~\ref{Gibbs} then gives
$\delta c_{aa}/c_{aa} = -2.1\pm$ 0.4 $\times10^{-4}$ at the SC $T_c$ of the overdoped
crystal rather than -1.2$\times10^{-4}$ obtained from Eq.~\ref{strain}. Importantly,
using Eq.~\ref{Gibbs} and  $\delta \alpha_a$ = -2.3$\times10^{-6}$/K, half of the
measured jump in $\delta \alpha_{b-a}$ for a fully oxygenated crystal~\cite{Meingast},
to obtain $\delta \mathbf{S}_{aa}$, gives $\delta c_{aa}/c_{aa}$ = -2.2$\times10^{-4}$
in good agreement with the value from $\delta\gamma$.

On the basis of the RUS data and  Eqs.~\ref{dilation},~\ref{strain} or \ref{Gibbs},
then assuming that $dT^*/dP$ and $dT_c/dP$ have similar values,  the changes in
$d\gamma(T)/dT $ and $d\alpha_a(T)/dT$ at $T^*$ should be $\approx$ 100 times larger
than the measured slopes of  $\gamma(T)$ and $\alpha_a(T)$ near 68~K, and should be
easily detectable. As Fig. 1 shows there is no discernible anomaly in $\gamma$ at
$T^*$ and similarly there is no
  clear anomaly in $\alpha_{b-a}$ near
  68~K for $\delta=1.0$~\cite{Meingast}.

Since these estimates of $\delta \gamma $ and $\delta \alpha_a$ from the mode
frequency changes are based on thermodynamic arguments they do not depend on the
detailed mechanism, but only assume  thermal equilibrium. The absence of associated
anomalies in $\gamma(T)$ and $\alpha(T)$ could, however, be explained if the anomalies
in $\delta f/f$ result, instead, from irreversible changes such as those associated
with anelastic relaxation, as we will see.

Evidence against a phase transition at 68~K for the over-doped crystal also emerges
when considering
 the proposed critical slowing down near $T^*$. The authors interpret the peak temperatures of the
 resonance widths at different frequencies in terms of coupling to fluctuations of the pseudogap order parameter.
  They ascribe this frequency dependence  to critical slowing down of these
   fluctuations  as $T \rightarrow T^*$. Modern theory of phase transitions tends to focus on the value of the
   dynamical exponents rather than magnitudes of physical parameters. However for electronic or magnetic phase
    transitions where quantum effects are important the time scale, $\tau$, of these fluctuations
    is often given by $\tau^{-1} \approx (k_B/h)|T - T^*|$ with a slope of $2.1 \times 10^{10}$Hz/K.
    Examples of this include time-dependent Ginzburg-Landau theory of superconductors \cite{Tinkham} and
    the critical slowing down of fluctuations in the antiferromagnet RbMnF$_4$ where
    the slope of $\tau^{-1}$ vs. $T$, near the N\'{e}el temperature of 83~K,
     is $\sim 1.2 \times 10^{10}$~Hz/K (see Fig. 9 of  Ref.~\cite{Halperin} based
     on neutron scattering data from Ref.~\cite{Tucciarone}). For YBCO with $T_c=88$~K
     Shekhter {\it et al.} observe (over a wide temperature range) a much smaller slope of $\sim 2 \times 10^5$~Hz/K which we feel
     is unlikely to be caused by critical fluctuations.

This low slope, combined with the absence of a relationship between $\delta f/f$ and
both  $\delta\gamma$ and $\delta\alpha_{b-a}$ at $T^*$, reinforces our view that the
features reported in Ref.~\cite{Shekhter} are more consistent with the effects of
anelastic relaxation. Indeed these features are reminiscent of those obtained in
earlier ultrasonic studies reviewed by Almond {\it et al.} \cite{Almond} where a small
activation energy $\sim$ 60 meV was
 found for one set of relaxation times. Fig. 2 shows the data from Fig.~4b in Ref.~\cite{Shekhter} interpreted
  as an inverse relaxation time, plotted on an Arrhenius plot. The data are parallel to those summarized by
   Almond {\it et al.}, they have a similar activation energy ($\sim$ 60 meV) but the attempt frequency is
   a factor of 10 lower. Such a low activation energy could, for example, be associated
   with hopping of copper and/or oxygen atoms between bistable sites \cite{Sullivan}.

\begin{figure}
\centerline{\includegraphics*[width=85mm]{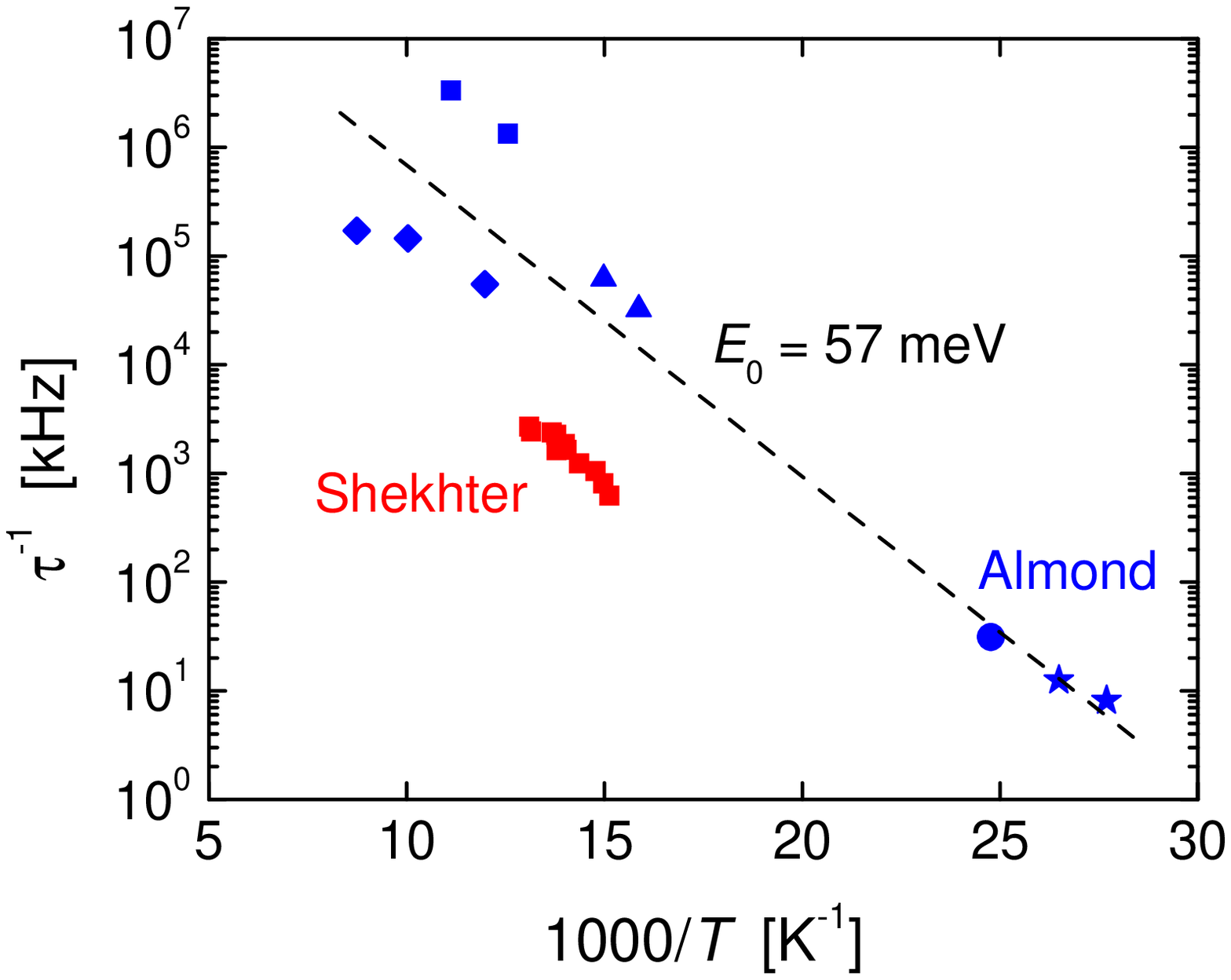}} \caption{\small (Color online)
An  Arrhenius plot of the data in Fig.~4b of Ref.~\cite{Shekhter} for the
YBa$_2$Cu$_3$O$_{6.98}$ crystal, where  $\tau^{-1} = 2\pi f$, $f$ is the RUS frequency
and
 $T$ is the temperature of the peak in RUS
width $\Gamma$. Also included are the data reported by Almond {\it et al.} from
ultrasonic anelastic relaxation studies. Both show an activation energy of about 60
meV.} \label{Arrhenius}
\end{figure}

We now turn to the anomalies near 245~K for the under-doped crystal shown in Fig.~2a
of Ref.~\cite{Shekhter}. In our own work we have found no anomalies or features in the
 specific heat in this temperature region that could not be explained by magnetic anomalies from a low concentration
  of CuO impurities. Early evidence for the sporadic appearance of anomalies in the $200-250$~K range was provisionally
  ascribed to ''a sluggish and
 hysteretic transition which may involve oxygen ordering" \cite{Junod}. This may still be a possibility in that  $245$~K is relatively
  close to the temperature of $280$~K where the measured magnetic susceptibility starts to depend on the cooling and
  warming rates \cite{Monod}. This hysteresis is typical of a kinetic transition involving oxygen disorder
   in the CuO chains, whose effects have also been observed in heat capacity and thermal expansion for oxygen
    deficient crystals above 280 K\cite{Nagel,Meingast2}.

 Alternatively, the abrupt slope changes, $d(\delta f/f)/dT =-1$ to $-2
\times 10^{-4}/$K below $T_0=245$~K in Fig.~2a of Ref.~\cite{Shekhter} can be ascribed
to  a second order phase transition smeared out over $\sim30$~K below 245~K or  to a
third order transition with $\delta F = a(T-T_0)^3$ and $a>0$. In either case the
slopes $d(\delta f/f)/dT$ and $d(\delta \gamma)/dT$ are related in the same way  as
$\delta f/f$ and $\delta \gamma$ in Eqs.~\ref{dilation},~\ref{strain} and less
directly~\ref{Gibbs}. If we assume that $T_0$ has the same moderate strain dependence
as $T_c$ of the overdoped SC crystal  then Eq.~\ref{strain} predicts an  increase of
0.5 to 1.1 mJ/mol/K$^3$ in the slope of $\gamma(T)$ below 245~K. These changes in
slope are equivalent to a change in $\gamma(T)$ over 30~K which is 0.6 to 1.2$\times$
the electronic term of fully oxygenated YBCO and  should have  been readily visible in
differential heat capacity measurements. They are actually a severe lower limit
because the reinforcement of
  the $a$ and $b$ axis contributions  in the strain-dependence of the SC $T_c$ described earlier is
  unlikely to apply to the strain-dependence of $T_0$
  for all three RUS frequencies reported.
But if $T_0$ were $\sim10 \times$ more strain-dependent than the SC $T_c$  then,
 for the same change in $df/dT$, Eq.~\ref{strain} gives a very small change in $d
\gamma/dT$ as observed. So to summarize,  we cannot exclude the possibility of a
highly strain-dependent phase transition at 245~K, but  the absence of a detectable
specific heat anomaly there clearly demonstrates (on entropy grounds) that the RUS
anomaly at 245~K cannot reflect the onset of the pseudogap.

As implied above, it is important to consider possible instabilities in the 100 to 300
K temperature region that could have a much larger effect on RUS data than on the
specific heat. The hard x-ray diffraction experiments
  for a YBCO crystal with $T_c = 67$~K \cite{Forgan1} reveal
   the $T$-dependence of the CDW amplitude below the CDW onset at 150~K, where any  oxygen-ordering effects
   are probably insignificant. Application of high
   magnetic fields shows that
the CDW and SC instabilities are in
   competition \cite{Forgan1,Julien1,Julien2} and therefore must have similar energy gaps.
Based on this, a model  calculation \cite{Cooper} suggests that if  a CDW with such a
$T$-dependent gap developed out of a metallic state with no pseudogap then it would
give a large anomaly in the heat capacity. However if the CDW onset
   occurs when the  DOS at the Fermi energy is already
  heavily depleted by the pseudogap then its effect on the heat capacity would be much less
  obvious. This is further evidence that  CDW or similar magnetic transitions
between 100 and 300~K  are \textit{not} causing the pseudogap.  RUS could be very
sensitive to these transitions since the mean field formula for the CDW transition
temperature
 of a quasi one-dimensional solid is similar to that for a weakly coupled
superconductor, but with the $\Theta_D$ pre-factor replaced by $T_F$, and can be  very
volume-dependent. However  the  volume-dependence of the pseudogap energy might also
play a part and could help clarify whether the CDW is caused by electron-lattice or
electron-electron interactions. Therefore  RUS experiments on an YBCO crystal with a
$T_c$ of 67~K, where the $T$-dependence of the CDW gap is known \cite{Forgan1}, could
give interesting results.

Irrespective of these  questions,  it not clear that the RUS data represent conclusive
evidence for counter-circulating current loops \cite{Zaanen}.  Some anomalous changes
in resonant frequency have been observed but their origin is undetermined and, at
least for the over-doped crystal, perhaps more consistent with thermally-activated
relaxation. Only one doping state (under-doped $T_c$ = 61.6~K) has a nominal $T^*$
which coincides with the neutron data. The over-doped sample ($T_c$ = 88~K) sits well
beyond the doping range of the neutron data.

In summary we maintain that the pseudogap $T^*$ line represents a crossover over a
broad temperature interval, as shown by the scaling behavior of the entropy,
susceptibility, resistivity, thermopower and Hall effect, over a wide range of
$T/T^*$. It reflects an underlying energy scale which falls to zero within the SC dome
and is definitely zero for well-oxygenated YBCO. There may be phase transitions near
the  experimentally-defined $T^*$ line but they are not having large effects on the
entropy of the
  charge carriers because they have not been seen in the heat capacity. It seems that they are not causing the
  pseudogap,
but instead they  involve charge or spin order developing in the pseudogap state.

  Funding from the Marsden Fund of New Zealand and the MacDiarmid Institute
 for Advanced Materials and Nanotechnology for JLT and JGS, from the EPSRC (UK) for JRC and JWL and from the
 Croatian Ministry of Science for IK is gratefully acknowledged.


\begin{references}
\bibitem{inversion} In this approximate treatment, $\mathbf{S}$ is obtained by
 inverting the  reduced, 3$\times$3  $c_{ij}$  matrix which contains only the non-shear elements  quoted in Ref.~\onlinecite{Welp}.
  Small changes in $\mathbf{S}_{ij}$ at $T_c$ given by Eq.~\ref{Gibbs}, are then made
under the plausible assumption that the Poisson ratio
$\mathbf{S}_{ab}/\mathbf{S}_{aa}$ does not change at $T_c$, and $\mathbf{S}$ is
inverted again to find the values of $c_{ij}$ below $T_c$. This gives reasonably good
agreement with Eq.~\ref{strain} but there  are discrepancies of almost a factor 2
which presumably arise from the the neglect of the unknown shear components of
$c_{ij}$.
\end{references}
\end{document}